\documentclass[preprint,12pt]{emulateapj-rtx4}
\usepackage[varg]{txfonts}
\usepackage{graphicx}
\usepackage{booktabs}
\usepackage{afterpage}
\usepackage{changepage}
\usepackage{chngcntr}
\bibpunct{(}{)}{;}{a}{}{,} 
\usepackage{color}
\usepackage[caption=false]{subfig}
\usepackage[colorlinks=true,citecolor=blue,urlcolor=blue]{hyperref}

\usepackage{longtable}

\newlength{\offsetpage}
{\end{adjustwidth}}

\begin{document}

\title{Orbital circularization of hot and cool Kepler eclipsing binaries}

\hyphenation{Kepler}

\author{
Vincent~Van~Eylen$^{1,2}$,
Joshua N.\ Winn$^{2,3}$,
Simon~Albrecht$^1$
%, Kevin Schlaufman$^2$
% and Saul Rappaport$^2$} 
}

\affil{
$^1$Stellar Astrophysics Centre, Department of Physics and Astronomy,
Aarhus University, Ny Munkegade 120, \\
DK-8000 Aarhus C, Denmark \\
$^2$MIT Kavli Institute for Astrophysics and Space Research, 70 Vassar St., Cambridge, MA 02139, USA \\
$^3$MIT Physics Department, 77 Mass.\ Ave.\ Cambridge, MA 02139, USA
}

\email{vincent@phys.au.dk}

\shorttitle{Circularization of hot/cool Kepler binaries}
\shortauthors{Van Eylen et al.}

\received{receipt date}
\revised{revision date}

\begin{abstract} {The rate of tidal circularization is predicted to be faster for relatively cool stars with convective outer layers, compared to hotter stars with radiative outer layers. Observing this effect is challenging, because it requires large and well-characterized samples including both hot and cool stars. Here we seek evidence for the predicted dependence of circularization upon stellar type, using a sample of 945 eclipsing binaries observed by \textit{Kepler}. This sample complements earlier studies of this effect, which employed smaller samples of better-characterized stars. For each \textit{Kepler} binary we measure $e \cos \omega$ based on the relative timing of the primary and secondary eclipses. We examine the distribution of $e\cos\omega$ as a function of period for binaries composed of hot stars, cool stars, and mixtures of the two types. At the shortest periods, hot-hot binaries are most likely to be eccentric; for periods shorter than 4 days, significant 
eccentricities occur frequently for hot-hot binaries, but not for hot-cool or cool-cool binaries. This is in qualitative agreement with theoretical expectations based on the slower dissipation rates of hot stars. However, the interpretation of our results is complicated by the largely unknown ages and evolutionary states of the stars in our sample.}
\end{abstract}


\maketitle
 
\section{Introduction}

Binary stars make up over half of all stars in the universe, and their orbital properties have been studied for many decades \citep[see, e.g.][]{kopal1956}. In close binary systems, tidal forces distort the shapes of stars and cause oscillations. The gradual dissipation of energy associated with those fluid motions ultimately leads to coplanarization and synchronization of rotational and orbital motion, as well as circularization of the binary orbit \citep[see, e.g.,][and references therein]{mazeh2008}.

One interesting aspect of tidal circularization theory is that the dissipation rate is a very strong function of the orbital semi-major axis, and thereby on the orbital period \citep[see, e.g.,][]{zahn1975}. Many efforts have been made to determine the so-called ``cut-off period'' \citep[see, e.g.,][]{mayor1984}, which characterizes the transition between mainly-circular and mainly-eccentric orbits. It has also been suggested that the cut-off period can serve as a proxy for age in star clusters \citep{mathieu1988}, and indeed there is evidence for a linear trend between the cut-off period and the age of binary stars in different clusters \citep[e.g.][]{meibom2005}.

Another interesting aspect of the theory is that the circularization timescale is predicted to depend strongly on stellar type. Stars with thick exterior convective zones are expected to experience more rapid tidal dissipation than stars with mainly-radiative exteriors \citep{zahn1975}.  Our interest in this topic was heightened by some recent developments in exoplanetary science.  The obliquities of the host stars of close-in giant planets have been observed to have different distributions for hot and cool stars \citep{winn2010,albrecht2012}, with the boundary between these types of stars at around $T_{\rm eff} \approx 6250$~K. It has been proposed that the differences in the obliquity distributions of cool and hot stars are due to differences in tidal dissipation, and perhaps magnetic braking \citep{dawson2014}.

Investigating these possibilities led us to search the literature for clear observational evidence of the dependence of tidal dissipation rates on effective temperature.  The literature does provide some evidence for the expected dependence of the cut-off period on stellar type, mainly through the comparison of different samples that have been analyzed in different ways. This is at least partly because cool stars and hot stars have been studied by different communities using different techniques. For hot stars, \cite{giuricin1984} found circularization below a period of 2 days for a sample of about 200 binary stars, which is seemingly consistent with the tidal friction theory of \cite{zahn1977}. A similar result was arrived at more recently by \cite{khaliullin2010}, who compiled a catalogue of about 100 eclipsing binaries from different sources. For cool stars, \cite{koch1981} investigated binaries with periods shorter than 20 days, and found reasonable agreement with theory. \cite{abt2006} collected 
eccentricities for cool stars and find cut-off periods at around 4 days.  \citet{pourbaix2004} maintain a large catalog of eclipsing binaries including both hot and cool stars, but the catalog is an inhomogeneous concatenation of various sources, and does not lend itself to statistical studies. The most convincing study to date was performed by \cite{torres2010}, who studied 95 eclipsing binaries with masses and radii known to better than 3\%. They found that short-period binaries tend to have circular orbits; long-period binaries have a wider range of eccentricities; and the critical period separating these regimes is larger for binaries with cool stars (defined by those authors as stars with $T_{\rm eff}< 7000$~K) than for binaries with hot stars.

Exoplanetary science has now provided an opportunity to perform a complementary study using a larger sample of less well-characterized binaries.
The NASA \textit{Kepler} mission \citep{borucki2010} was designed to find transiting planets, and also discovered thousands of eclipsing binaries \citep{prsa2011,slawson2011}. The \textit{Kepler} photometry can be used to precisely determine one component of orbital eccentricity: $e\cos\omega$, the product of the eccentricity and the cosine of the argument of pericenter, which is related to the relative timing of the primary and secondary eclipses.  The measured eclipse durations can also be used to determine $e\sin\omega$, though this is less straightforward. These photometric constraints on eccentricity are much easier to obtain than the task of measuring the orbital eccentricity through radial-velocity monitoring \citep[see, e.g.,][]{mazeh2006}. Although \cite{slawson2011} modeled a large number of \textit{Kepler} EBs, the results for eccentricity were not reliable, 
presumably because their neural-network approach 
was designed to measure many different properties for a wide variety of binary stars and was not trained specifically to determine eccentricities.

Here, we employ a simpler method to measure $e\cos\omega$ reliably, based only on the relative timing of the primary and secondary eclipses. We combine these measurements with published effective temperatures for the binary components \citep{armstrong2014} to divide our sample into hot-hot, hot-cool, and cool-cool binaries. We then compare the observed $e\cos\omega$--period diagrams for these different categories of EBs. Thus, our study benefits from a relatively large sample and from homogeneity in the measurement techniques. However, it has the significant drawback that the masses, sizes, ages, and evolutionary states of the stars are not nearly as well known as the systems studied by \citet{torres2010}.

This paper is organized as follows. Section~\ref{sec:methods} describes our sample and measurement techniques. Section~\ref{sec:results} presents the results. Section~\ref{sec:theory} compares our results to simple theoretical expectations of tidal dissipation rates, taking into account systematic differences in stellar age. Finally, a discussion is presented in Section~\ref{sec:discussion}.

\section{Methods}
\label{sec:methods}

\subsection{Measuring $e \cos \omega$}

To create our sample we begin with the EB catalog by \cite{slawson2011}\footnote{We use an updated version that is available online at {\url{http://keplerebs.villanova.edu/}}, accessed on 9 March 2015.}. We use the periods given in this catalog. We rely on the effective temperatures for the primary and secondary stars, $T_1$ and $T_2$, that were determined by \citet{armstrong2014} by fitting the observed spectral energy distribution.  We impose the restriction $T_\mathrm{eff} \leq 10,000~K$, to limit the number of very young stars in the sample and thereby simplify the interpretation, as described in Section~\ref{sec:theory} in more detail. To measure $e \cos \omega$, we determine the times of primary and secondary eclipses, using the procedure described below.

The \textit{Kepler} observations are separated into different quarters (Q), each representing about three months of data. The data from each quarter are provided in three separate files, each containing one month of data. Our starting point is the pre-search data conditioning (PDC) photometry, from which some of the instrumental trends have been removed \citep{smith2012}.  For normalization, we divide the flux data from each month by the mean monthly flux level.  The data are then folded based on the period reported in the Villanova EB catalog \citep{slawson2011}, and binned in orbital phase by a factor of 50.

Subsequently we determine the times of the primary and secondary eclipses. We do so as follows: first, we locate the approximate time of the deepest eclipse ($t_1$). The time interval containing this eclipse is then ignored, and we determine the approximate time of the second eclipse ($t_2$). We determine the precise times by fitting a second-order polynomial function of time to the data near minimum light. We estimate uncertainties by using a bootstrap technique. We create 100 samples by drawing randomly (with repetitions allowed) from the actual data points. We then refit these samples, and adopt the mean of the results as the ``best value'', and the standard deviation of the results as the 1$\sigma$ uncertainty.

\begin{figure}[h!]
\centering
\includegraphics[width=\linewidth]{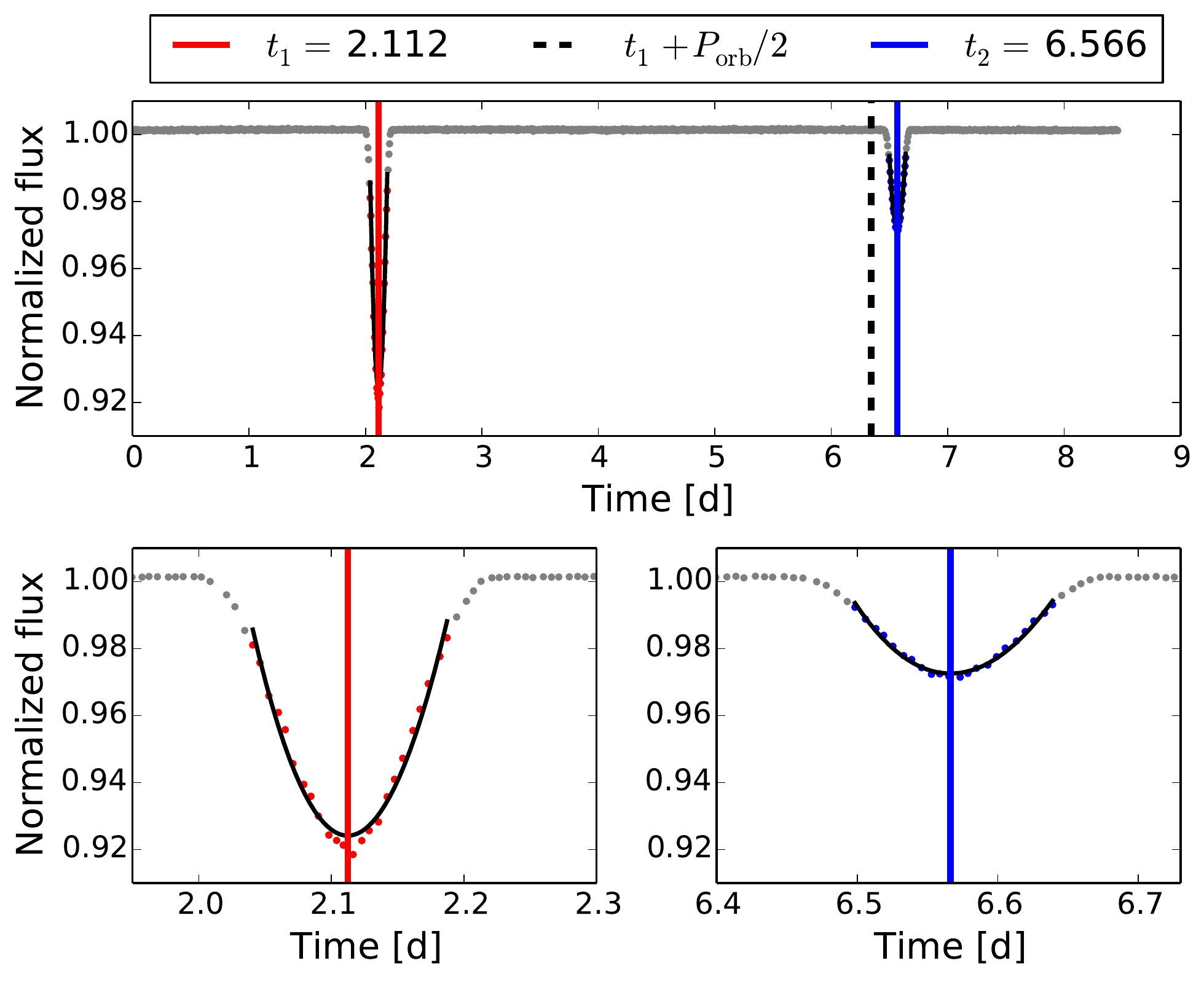} 
\caption[]{Sample output from our code for
finding primary and secondary eclipse times. The colored data points are fitted to determine the eclipse times; the results are marked by thick vertical lines. The dotted vertical line indicates the expected location of the secondary eclipse, for a circular orbit. The bottom panels allow closer inspection of the narrow time ranges surrounding the primary and secondary eclipses.}
\label{fig:transittimefinder}
\end{figure}

We then find $e \cos \omega$, using
\begin{equation}
e \cos \omega \approx \frac{\pi (t_2 - t_1)}{2P} - \pi/4
\end{equation}
where $P$ is the orbital period.\footnote{Note that this equation is only approximate and correction factors apply for high eccentricity; see e.g.\ \cite{sterne1940}.} The method is illustrated in Figure~\ref{fig:transittimefinder}. 

%%%%%%%%%%%%%%%%%%%%%%%%
%%%%%%%%%%%%%%%%%%%%%%%%

We apply this method to all stars in the EB catalog \citep{slawson2011} which have orbital periods between 1.5--50 days. Binary systems with periods shorter than 1.5 days are often non-detached, which complicates the measurements.
These binaries are not likely to be useful for our study because they have been found to be circularized regardless of stellar type \citep{torres2010}. Likewise, for binaries with periods longer than 50 days, tidal circularization is very likely 
irrelevant for main-sequence binaries of all spectral types.
%\enlargethispage{1cm}

In some cases, our automated method fails to identify the correct eclipse times due to data artifacts. These systems are manually refitted. To avoid any bias, we perform this refitting blindly, i.e., the fitter has no knowledge of the stellar effective temperatures. In other cases there are no secondary eclipses, or no primary eclipses. We remove these cases from consideration, after confirming the absence of the eclipses through visual inspection of all the folded light curves. We do not think that these omissions produce any significant bias relating to stellar effective temperature, although we caution that highly eccentric binaries are more likely to only show either primary or secondary eclipses, which means that our derived distribution of $e\cos\omega$ (for all stellar temperatures) may be biased towards lower values.

\section{Results}
\label{sec:results}

\begin{figure}%[h!]
\centering
\includegraphics[width=\linewidth]{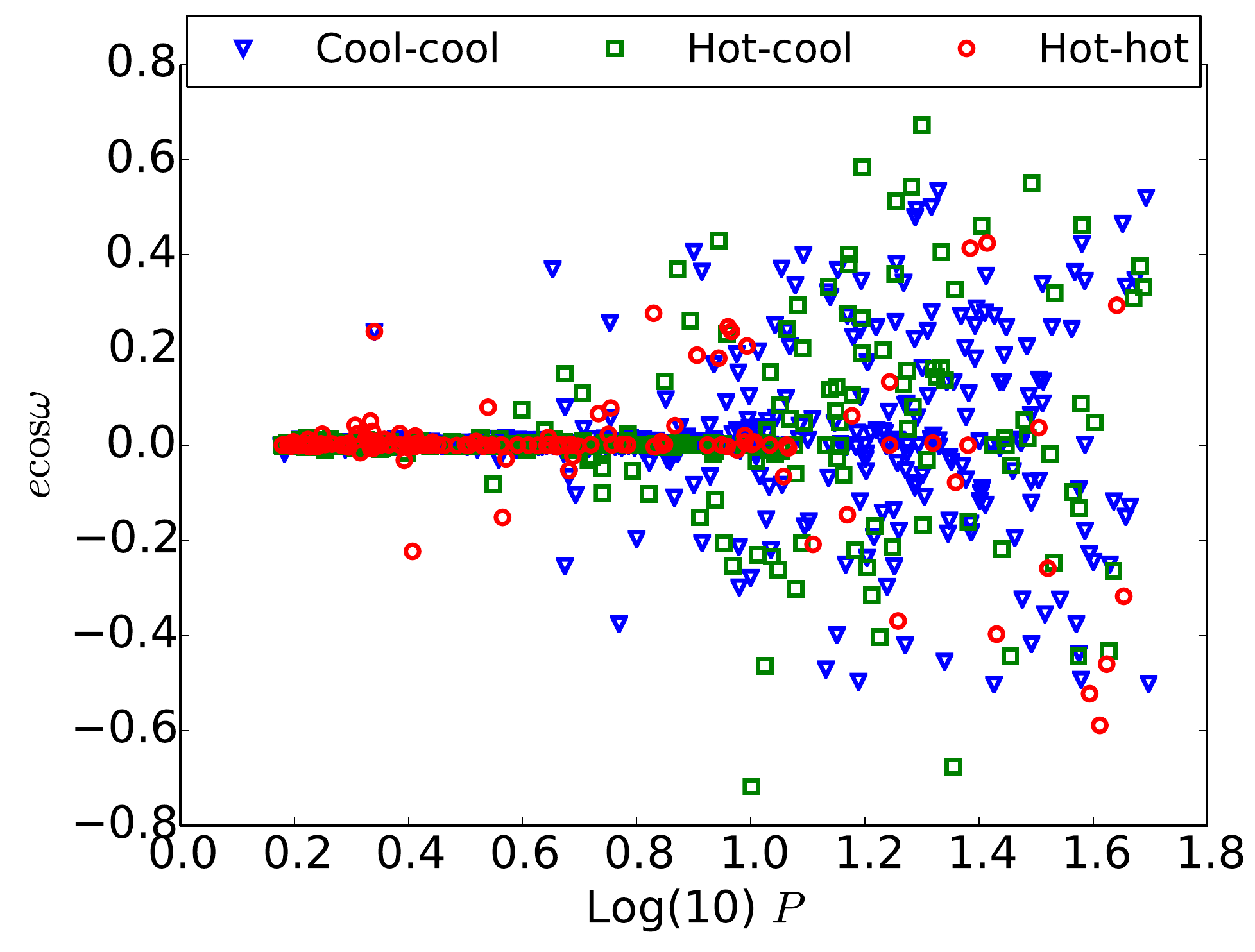} 
\caption[]{Measurements of $e \cos \omega$ as function of orbital period. The
red circles represent ``hot-hot'' binaries in which both stars have
$T_{\rm eff}>6250$~K.
Blue triangles are for ``cool-cool'' binaries, and green squares
are for binaries with one hot star and one cool star.
%The lines illustrate the cut-off period for the different categories, using Equation~\ref{eq:cutoff}, and are fitted to $|e\cos\omega|$.
}
\label{fig:periodecc_scatter}
\end{figure}

The main result of our analysis are measurements of $e\cos\omega$ that are associated with the previously determined orbital period $P$ \citep[from][]{slawson2011} and the stellar temperatures of the two components \citep[from][]{armstrong2014}. Figure~\ref{fig:periodecc_scatter} shows the measurements of $e \cos \omega$ as a function of $P$, with the color of each circle encoding the effective temperatures of the components. Following \cite{winn2010} and \cite{albrecht2012}, we use 6250~K as the boundary between ``hot'' and ``cool'' stars. A binary is designated ``hot-hot'' if both stars have estimated effective temperatures exceeding this nominal boundary value. Likewise, ``cool-cool'' binaries have two cooler stars, and ``hot-cool'' binaries have one hot star and one cool star.  We present results for 137 hot-hot binaries, 289 hot-cool binaries and 519 cool-cool binaries. The parameters for all EBs in our sample are reported in Table~\ref{tab:parameters}.

%%%%%%%%%%%%%%%%%%%%%%%%
%%%%%%%%%%%%%%%%%%%%%%%%

\begin{table*}
  \begin{center}
    \caption{System parameters for \textit{Kepler} eclipsing binaries with measured $e \cos \omega$ values. \label{tab:parameters}}
    \smallskip
    \begin{tabular}{ccccccccccccccc}
KIC & Period [d] & Ephemeris [HJD] & T$_1$ [K] & $\sigma$ (T$_1$) [K] & T$_2$ [K] & $\sigma$ (T$_2$) [K] & $e \cos \omega$ & $\sigma (e \cos \omega)$ \\ 
\hline 
1026032&	8.46044&	54966.77381&	6149&	359&	4863&	556&	0.0416&	0.0005\\
1433980&	1.59263&	55000.37409&	6869&	370&	5484&	558&	-0.0014&	0.0021\\
1571511&	14.02245&	54954.50475&	5946&	363&	6023&	684&	0.0465&	0.0021\\
1575690&	2.25243&	54965.43526&	4207&	370&	3776&	570&	-0.0015&	0.0008\\
1725193&	5.92769&	55005.64981&	6044&	354&	5988&	554&	0.0002&	0.0008\\
2019076&	7.12923&	55004.07222&	6199&	360&	5184&	561&	0.0&	0.0008\\
2161623&	2.28347&	54999.59984&	7045&	765&	5106&	1085&	0.0049&	0.0048\\
2162994&	4.10159&	54965.63165&	5823&	357&	5684&	551&	0.0003&	0.0007\\
2167890&	2.6483&	55185.84325&	4878&	370&	4765&	623&	0.008&	0.0074\\
2306740&	10.30699&	54966.42521&	6060&	368&	5769&	656&	0.0238&	0.0006\\
2308957&	2.21968&	54965.16646&	5993&	363&	5825&	562&	0.0016&	0.001\\
2309587&	1.83851&	54965.10135&	5765&	381&	5576&	773&	0.0019&	0.0013\\
2437060&	3.18711&	55000.89072&	4930&	357&	3860&	576&	0.0027&	0.0096\\
2437149&	18.79874&	55008.62182&	5456&	361&	5165&	624&	-0.0165&	0.0022\\
2437452&	14.46993&	54974.81653&	5488&	360&	4552&	609&	0.0012&	0.0005\\
2437783&	7.45341&	55002.00678&	6149&	1599&	6343&	2474&	0.0042&	0.0041\\
2438061&	4.88585&	55004.09363&	5278&	356&	4559&	567&	-0.0001&	0.0009\\
2438490&	3.31577&	55001.02005&	5459&	376&	5220&	683&	-0.009&	0.0061\\
2441161&	4.38398&	55004.68452&	5501&	362&	4887&	570&	-0.0016&	0.0007\\
2442084&	49.7886&	55008.19209&	3970&	352&	3819&	553&	-0.5019&	0.0014\\
% 2445134&	8.41201&	54972.64775&	6467&	355&	3958&	692&	0.0009&	0.0005\\
% 2449090&	4.944&	55001.81104&	5690&	405&	4284&	712&	-0.0011&	0.0013\\
% 2450566&	1.84459&	55001.5801&	6711&	359&	6775&	596&	0.0024&	0.0007\\
% 2452440&	8.09689&	55003.83758&	5801&	688&	5516&	1193&	0.0015&	0.0056\\
% 2569494&	1.52332&	55000.34919&	5156&	500&	4856&	906&	-0.0178&	0.0055\\
% 2580872&	15.92663&	54978.5513&	5267&	383&	4878&	704&	-0.0545&	0.0005\\
% 2693092&	39.84152&	54961.29894&	5926&	353&	4129&	953&	-0.2455&	0.0005\\
% 2695740&	3.61591&	54956.42553&	5268&	761&	4501&	1078&	-0.0305&	0.0074\\
% 2708156&	1.89127&	54954.33559&	15058&	515&	10410&	2484&	-0.0014&	0.0015\\
\multicolumn{8}{l}{Notes:}\\
\multicolumn{8}{l}{1. Full version of the table available online.}
\end{tabular}
\end{center}
\end{table*}

Figure~\ref{fig:periodecc_scatter} shows that the distribution of $e\cos\omega$ is roughly symmetric around zero, as would be expected for a uniform random distribution of $\omega$, the argument of pericenter.  At the shortest periods, it is clear that most binaries are circular or nearly circular. The spread in eccentricity increases with increasing orbital period. This is consistent with what is expected from tidal circularization at short periods, and with previous studies
of other samples.

%%%%%%%%%%%%%%%%%%%%%%%%
\begin{figure}%[h!]
\centering
\includegraphics[width=	 \linewidth]{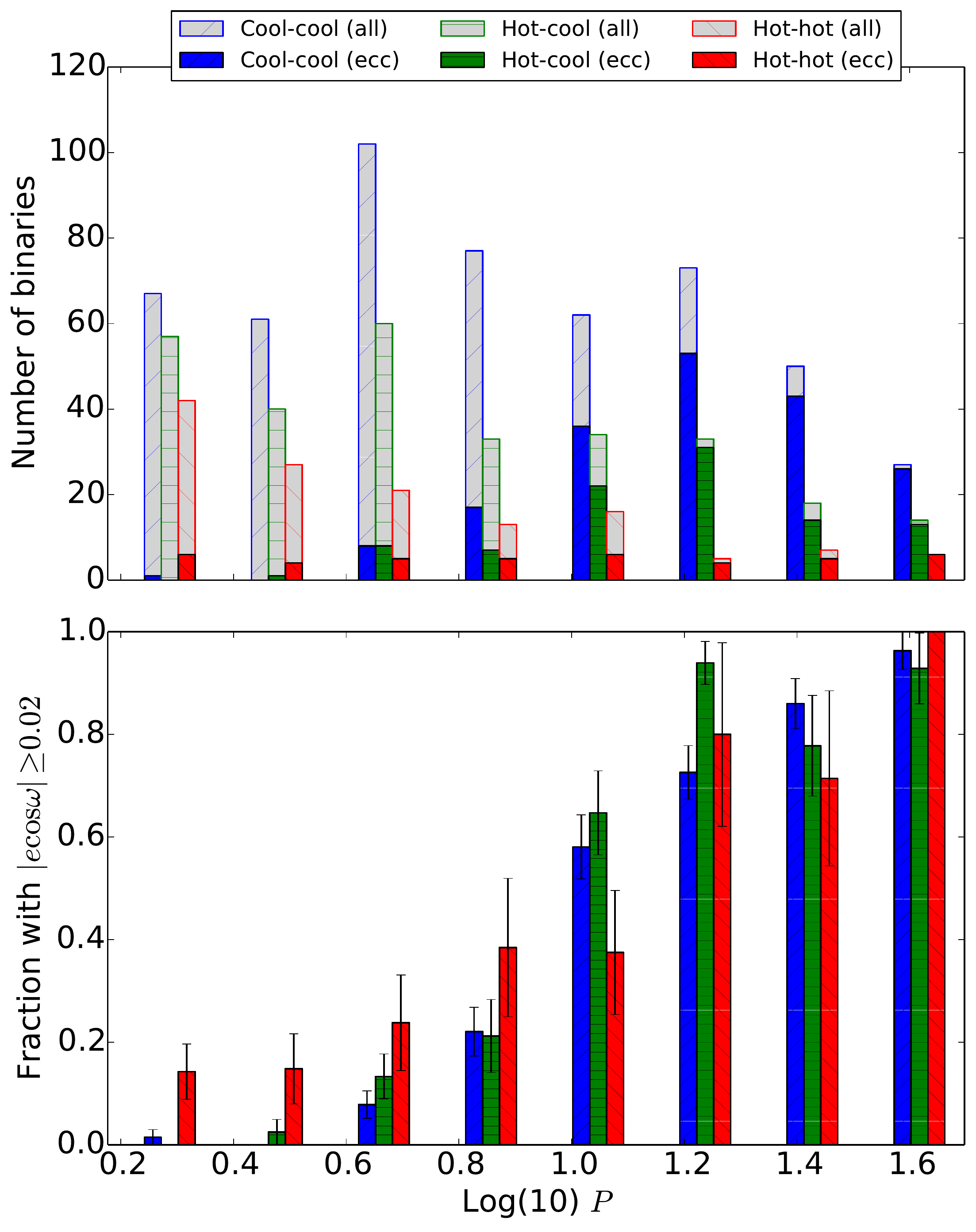} 
\caption[]{\textit{Top:} histogram of the binaries in (logarithmic) period bins, divided in hot-hot, hot-cool, and cool-cool binaries. The eccentric binaries, defined as $|e\cos\omega|~\geq~0.02$, are highlighted. \textit{Bottom:} the fraction of eccentric binaries per bin.}
\label{fig:period_vs_fraction}
\end{figure}
%%%%%%%%%%%%%%%%%%%%%%%%

Figure~\ref{fig:period_vs_fraction} shows the fraction of EBs that are significantly eccentric, within different period bins.  In this analysis, we use
$|e \cos \omega| \geq 0.02$ as our criterion for
significant eccentricity. The error bars displayed on the bins represent only the uncertainty due to Poisson (counting) statistics. It is clear that beyond 10-15 days, the large majority of binaries are eccentric regardless of the temperature. At shorter periods, this is not the case. The fraction of eccentric binaries decreases with decreasing period, and it does so at different rates for different temperature classes.
For $P\leq 10$~days, the hot-hot binaries have a greater fraction of eccentric systems at a given period.

\section{Comparison with theory}
\label{sec:theory}
The observed eccentricity of an EB depends on its initial eccentricity, the rate of tidal circularization, and the time interval over which circularization has taken place, i.e., the age of the system. For simplicity we assume that the initial eccentricity distribution is broad and is the same for binaries of all types (although we are not aware of any firm observational support for this latter assumption). Thus, in our model, any differences in eccentricity distributions between samples of EBs come from differences in tidal dissipation rates and ages.

To calculate the expected timescales for tidal circularization for convective stars ($\tau_\textrm{conv}$) and for radiative stars ($\tau_\textrm{rad}$), we follow
\cite{claret1997} and use the formulas
\begin{equation}
 \tau_\textrm{circ,~conv} =
(1.99 \times 10^3~{\rm yr})
M^3 \frac{(1+q)^{5/3}}{q} L^{-1/3} \lambda_2^{-1} \frac{P^{16/3}}{R^{22/3}} \label{eq:timeconvection}
\end{equation}
and 
\begin{equation}
 \tau_\textrm{circ,~rad} = (1.71 \times 10^1~{\rm yr})
M^3 \frac{(1+q)^{5/3}}{q} E_2^{-1} \frac{P^7}{R^9},
\label{eq:timeradiation}
\end{equation}
where $M$ and $R$ are the total stellar mass and radius in solar units, $q$ is the mass ratio of the two stars, and $L$ is the total luminosity relative to the Sun's luminosity. For simplicity we assume the main-sequence relations $L \propto M^{3.9}$ and $R \propto M^{0.8}$, and we set $q = 1$. We use representative values for $E_2$ and $\lambda_2$ \cite[][see their Figures 1 and 3]{claret1997}. We can now calculate the circularization timescale for convective and radiative stars of different masses and periods.

As for the age, our estimate is based on the simple and approximate main-sequence relationship, $\tau_\star = (10^{10}~{\rm yr}) (M/M_\odot)^{-2.9}$.  While ages of individual \textit{Kepler} EBs are typically unknown, on average the hot stars are expected to be systematically more massive, faster-evolving, and younger, giving them less time for tidal dissipation to circularize their orbits.

%%%%%%%%%%%%%%%%%%%%%%%%
\begin{figure}[th!]
\centering
\includegraphics[width=\linewidth]{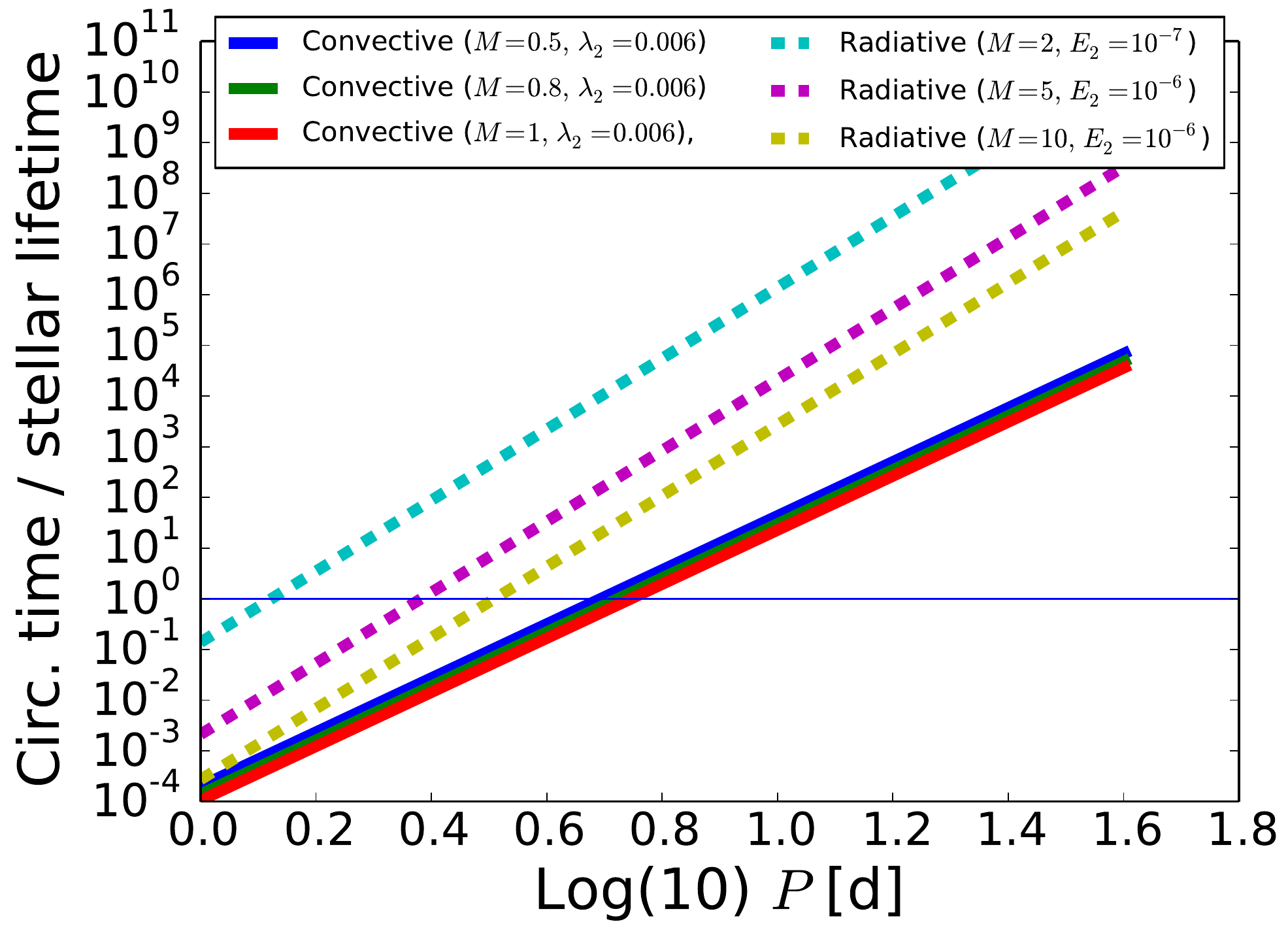} 
\includegraphics[width=\linewidth]{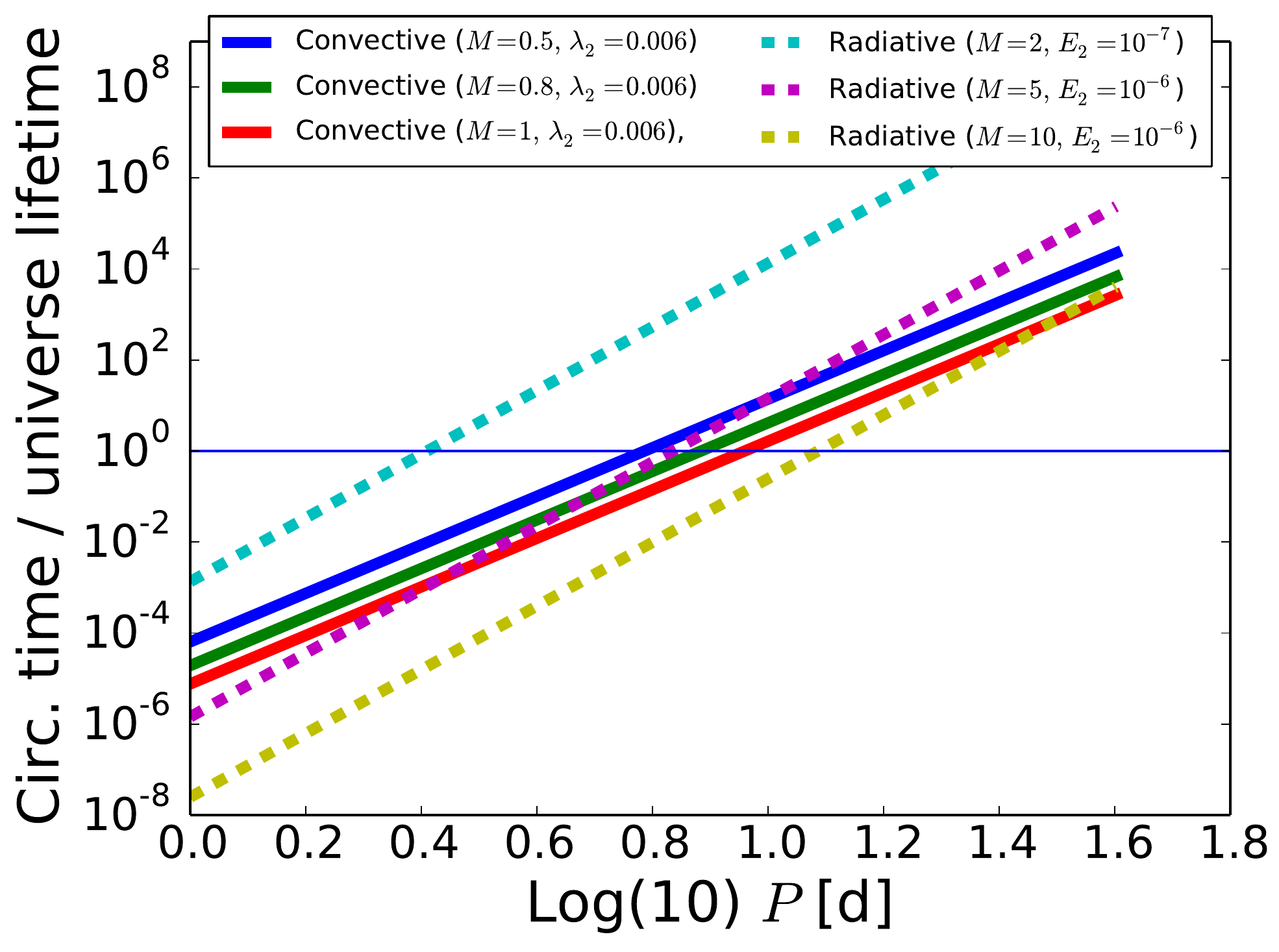} 
\caption[]{\textbf{Top:} The circularization time scales for convective (Equation~\ref{eq:timeconvection}) and radiative (Equation~\ref{eq:timeradiation}) stars, divided by the typical lifetime of such a star. Numbers above unity indicate that most of these systems haven't had time to circularize, while numbers below unity indicate a circularization timescale shorter than the lifetime of the system. \textbf{Bottom:} Rather than dividing by the stellar lifetime, we have now divided by the age of the universe. If the most massive stars would live this long, they would circularize up to longer periods than they do, but stars of a few solar masses are much less affected.}
\label{fig:toymodel}
\end{figure}
%%%%%%%%%%%%%%%%%%%%%%%%

We can now calculate the relative circularization time $\tau_{\rm circ}/\tau_\star$ as a function of period for different systems. Figure~\ref{fig:toymodel} shows some illustrative results.  For values of $\tau_{\rm circ}/\tau_\star$ greatly exceeding unity, we expect to observe a broad distribution of eccentricities because the lifetime of the system is too short to have allowed for significant circularization.  For values of $\tau_{\rm circ}/\tau_\star$ well below unity, the opposite is true, and we expect most binaries to have circular orbits.

According to the results of this rough calculation, we should expect to find a critical circularization period for cool stars in the neighborhood of 5~days ($\log P = 0.7$). For hot stars we should find a shorter circularization period depending more strongly on mass, ranging from about 1.3~days for 10~$M_\odot$ to 3~days for 2~$M_\odot$. We emphasize that these exact numbers depend on a range of assumptions outlined above, and should be considered as rough estimates rather than exact predictions.

To isolate the theoretical effect of tidal dissipation, as opposed to stellar age, Figure~\ref{fig:toymodel} shows the results after setting $\tau_\star$ equal to the Hubble time for all cases. Here there are no relative age effects at play. By comparing the two panels in this figure we see that the effect of age is important for the most massive stars (which are on average the youngest).  For the hot stars, replacing the stellar age with the much longer Hubble time results in longer circularization periods. Thus, in this simple picture, stars with a mass of $5~M_\odot$ have a similar tidal dissipation rate as cooler, lower-mass stars with thick convective zones, but we should nevertheless observe such stars to have a shorter circularization period because of their younger ages.  For even more massive stars the age effect becomes even more dominant. For the case of 10~$M_\odot$, the circularization period actually {\it exceeds} that of the cooler stars.  On the other hand, for stars of mass $2~M_\odot$, age 
effects are much less important. This is understandable, since the typical lifetime of such stars is only a few times shorter than that of subsolar mass stars.

These theoretical considerations show that for the purpose of observing the specific dependence of tidal dissipation rates on stellar type, it is important to focus on stars that are not too massive (too hot). We are thereby led to limit our sample to $M \le 3~M_\odot$.  Since $T_{\rm eff} / T_{{\rm eff},\odot} \approx \sqrt{M/M_\odot}$ on the main sequence, this is achieved by restricting $T_\mathrm{eff} \leq 10,000~K$.

Based on Figure~\ref{fig:toymodel} we expected that cool-cool binaries would be mainly circular below $\approx$5~days ($\log P = 0.7$), with the precise ``cut-off'' period almost independent of stellar mass.  For hot-hot binaries we expected that for periods exceeding 2-3 days ($\log P = 0.3-0.5$), some systems could remain eccentric, with the precise cut-off period depending more strongly on stellar mass, leading to a greater scatter. For hot-cool systems, the cool component is expected to provide most of the tidal dissipation, leading us to expect the cut-off period for hot-cool systems to be similar to that of cool-cool systems.\footnote{In reality the situation may be more complex if the hot component is significantly more massive, because the total mass affects the orbital separation at a given orbital period. We have not attempted to correct for this effect given the coarse knowledge of stellar masses in our sample.}

\begin{figure}[ht]
\centering
\includegraphics[width=\linewidth]{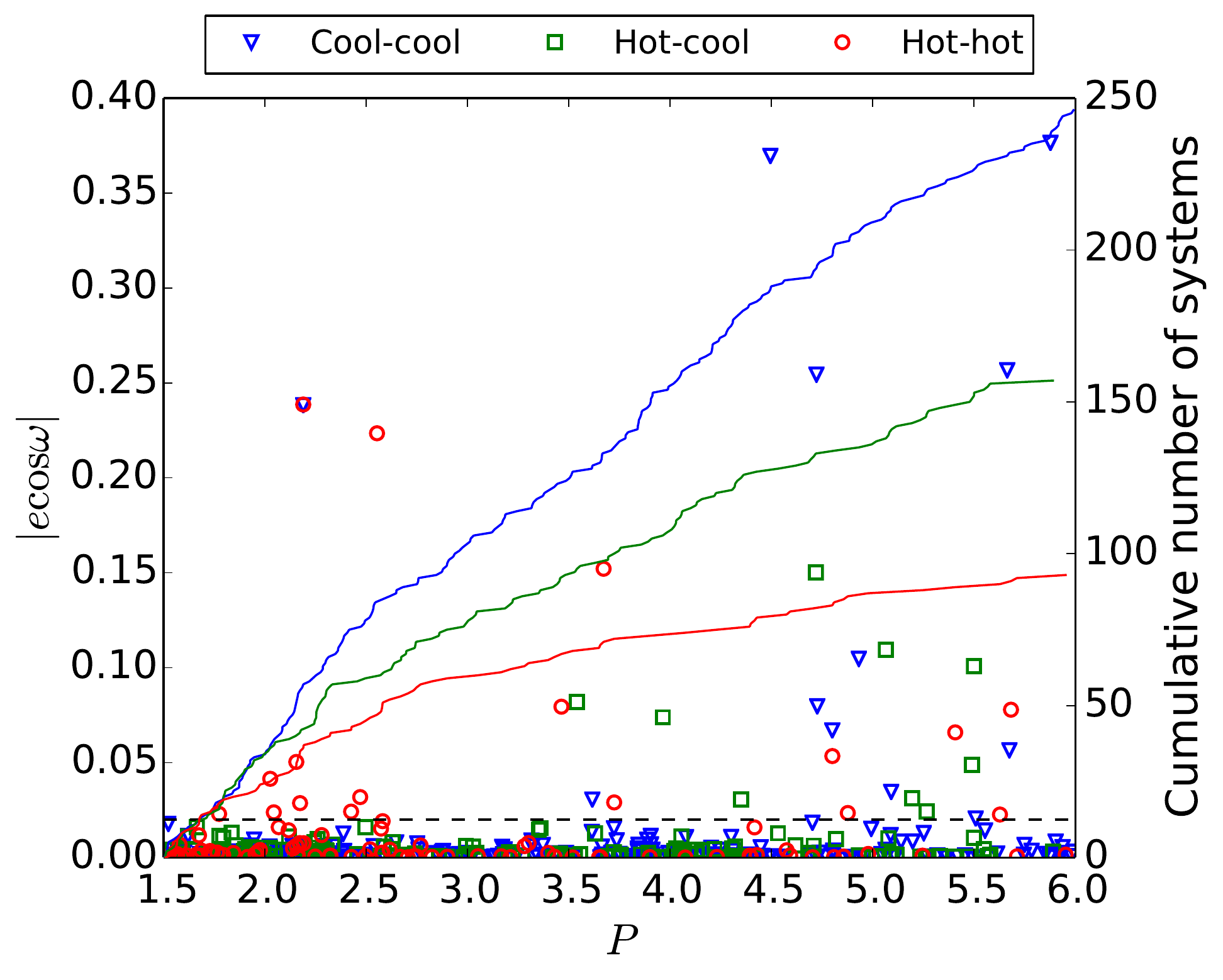} 
\caption[]{Measurements of $|e \cos \omega|$ as function of orbital period in the period range of 2-6 days. The red circles represent ``hot-hot'' binaries in which both stars have $T_{\rm eff}>6250$~K.
Blue triangles are for ``cool-cool'' binaries, and green squares are for binaries with one hot star and one cool star. Despite the smaller number of hot-hot binaries, this category boasts the largest number of eccentric systems.\\
}
\label{fig:ecosomega_period_zoom}
\end{figure}

Qualitatively, these are indeed the trends that are observed.
Figure~\ref{fig:ecosomega_period_zoom} allows a closer inspection of
the systems with periods shorter than 5 days.
At the shortest periods, a clear difference is observed between stars with different temperatures. For example, at periods between 1.5 and 4 days, there are 74 hot-hot stars of which 12 show significant eccentricity (which, as before, is defined as $|e \cos \omega| \geq 0.02$). By contrast, there are 156 cool-cool and 108 hot-cool stars, and only 2 stars in each category have a significant eccentricity. For periods between 4 and 5 days, some eccentric systems occur for all stellar types: there are 2 out of 14 hot-hot systems showing eccentricity, 5 out of 54 for cool-cool systems, and 2 out of 29 for hot-cool systems.

\begin{figure}[ht]
\centering
\includegraphics[width=\linewidth]{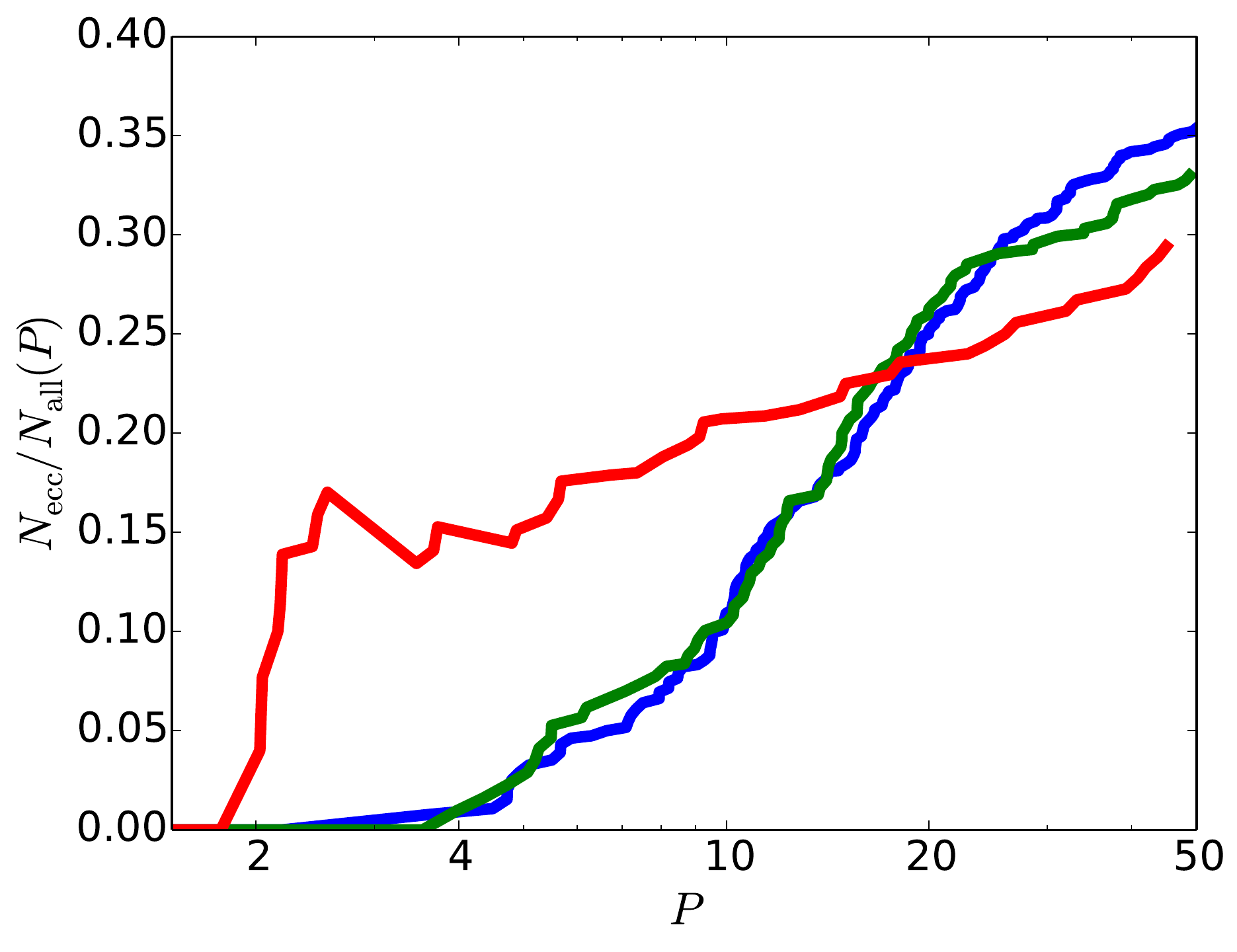} 
\caption[]{Cumulative fraction of eccentric systems, as defined in Equation~\ref{eq:fractional_cumulative}. Hot-hot binaries can be eccentric at periods beyond about 2 days, while hot-cool and cool-cool binaries exhibit eccentricities at periods longer than 4-5 days.
}
\label{fig:fractional_cumulative}
\end{figure}

In Figure~\ref{fig:fractional_cumulative}, we show the cumulative fraction of systems with a significant eccentricity, $N_\mathrm{ecc}/N_\mathrm{all}(P)$. This is defined as
\begin{equation}
\label{eq:fractional_cumulative}
\frac{N_\mathrm{ecc}}{N_\mathrm{all}}(P) =  \frac{N {(P_\star \leq P \mathrm{~and~} |e \cos \omega| \geq 0.02)}} {N {(P_\star \leq P)}}
\end{equation}
where $N(P_\star \leq P)$ is the number of binary stars with periods less than or equal to $P$. From Figure~\ref{fig:fractional_cumulative}, we see that hot-hot binaries show significant eccentricities at shorter periods than hot-cool or cool-cool binaries. The hot-cool and cool-cool binaries have a very similar period-dependence of their eccentricity distributions. At longer periods, the hot-cool and cool-cool binaries have a higher fraction of eccentric systems than the hot-hot binaries, presumably because the sample of hot-hot binaries has a larger fraction of shorter-period systems (see Figure~\ref{fig:period_vs_fraction}).

%%%%%%%%%%%%%%%%%%%%%%%%
%%%%%%%%%%%%%%%%%%%%%%%%

%%%%%%%%%%%%%%%%%%%%%%%%
%%%%%%%%%%%%%%%%%%%%%%%%

\section{Discussion}
\label{sec:discussion}

This work represents an attempt to compare the degree to which tidal circularization has taken place in binaries with hot and cool stars, using a large sample of \textit{Kepler} eclipsing binaries that has been analyzed with homogeneous measurement techniques.  The observations displayed in Figure~\ref{fig:periodecc_scatter} and Figure~\ref{fig:period_vs_fraction} agree remarkably well with the predictions in Figure~\ref{fig:toymodel}, despite the obvious limitations and simplicity of our crude theoretical calculations.

This suggests that we have indeed detected the dependence of orbital circularization on stellar type, due to the combination of age effects and differing tidal dissipation rates. The observations alone cannot tell us which of these two factors --- age or tidal dissipation --- is more important. However, as pointed out in Section~\ref{sec:theory}, for stars cooler than about $10,000~$K, age effects are not expected to be dominant; the main effect should be tidal dissipation rates.  Another suggestion that the differences in cut-off periods between the samples cannot be caused exclusively by differences in age is provided by the ``mixed'' binaries. In these cases one of the stars is hot and therefore evolving rapidly, causing these binary systems to be systematically younger than cool-cool binaries. Thus, in terms of age, the mixed binaries should be comparable to the hot-hot systems. However, we observe their eccentricity fraction at short periods to be lower than that of the hot-hot systems, which (in our 
admittedly simple framework) can only be explained by the higher tidal dissipation rate of the cool component in these systems.

Despite the advantages of a homogeneous analysis method and relatively large sample, there are also important limitations of our study. Rather than eccentricity itself, we chose to focus on $e\cos\omega$, because this parameter is so readily determined from the existing \textit{Kepler} photometry. After some preliminary efforts, we abandoned the attempt to measure $e\sin\omega$, which can in principle be derived from the relative duration of the primary and secondary eclipses.  We found that such measurements are considerably more complicated to make reliably, due to the covariance of this parameter with the semi-major axis, orbital inclination, and limb-darkening parameters.  Even in favorable cases the precision in $e\sin\omega$ is typically an order of magnitude worse than in $e\cos\omega$. Nevertheless, it is likely that the measurements of both $e\cos\omega$ and $e\sin\omega$ can be improved upon for individual systems of interest. In addition, radial-velocity observations could be undertaken to 
validate these determinations.

In this work we have taken 6250~K to be the dividing line between hot stars with radiative outer layers and cool stars with convective outer layers. This choice was made for consistency with previous work on the obliquities of exoplanet host stars \citep{winn2010,albrecht2012}. However, other dividing lines have been used by other authors; for example, \cite{torres2010} used 7000~K. If we use 7000~K as the dividing line, there are 8 hot-hot systems out of 34 with periods shorter than 4 days which exhibit a significant eccentricity (as opposed to 12 out of 74, for a dividing line of
6250~K). Furthermore, with the 7000~K boundary, there would be 6 out of 259 cool-cool eccentric systems, and 2 out of 45 hot-cool eccentric systems. This suggests that if we had chosen 7000~K, this would have strengthened the result that hot-hot binaries are
more likely than hot-cool or cool-cool binaries to be eccentric at relatively short periods. We also note that \cite{torres2010} did not impose an upper limit of $10,000~$K, as we do here. A substantial fraction of the hot stars in their sample are substantially hotter than those considered here (the hottest stars in their sample is $38,000~$K). As a result, the \cite{torres2010} sample may be more affected by the systematic age differences between hot and cool stars.

We did not take into account the uncertainties in the stellar temperatures given by \cite{armstrong2014}. In some cases the uncertainties are substantial: the mean uncertainty for primary stars is 370~K. Undoubtedly some of the objects in our sample have been misclassified. This situation will improve after the EBs are studied spectroscopically. We also note that the calculations of $T_2$ relative to $T_1$ by \cite{armstrong2014} may have a subtle dependence on eccentricity, because
the $e\sin \omega$ parameter can affect the depth ratio between eclipses. We neglected this effect in this study.

There are also some issues to keep in mind regarding the theoretical interpretation. As pointed out in Section~\ref{sec:theory}, we have assumed the initial eccentricity distribution is the same for all stellar types, which is not necessarily the case. In addition, we do not generally know the age of individual binary stars. At this point we can only make general statements about systematic differences between our samples. In some cases the individual system ages could be derived from isochrone fitting, asteroseismology or gyrochronology, although deriving ages for all EBs in the sample would require a considerable effort. Furthermore, in our interpretation we have assumed that all stars in our sample are on the main-sequence, by using the scaling relations presented in Section~\ref{sec:theory}. This is certainly not the case in reality, particularly for the hotter and faster-evolving stars. We further assumed that all binaries with periods longer than 1.5 days are detached, but in reality some of 
them may be semi-detached.

We have attributed the differences in the eccentricity distribution, in part, to tidal effects, using simplified equations drawn from equilibrium and dynamical tidal circularization theory, as brought forward by \cite{zahn1975}. In reality, the tidal circularization efficiency is probably itself a function of stellar evolution, i.e.\ it is not necessarily the same throughout the evolution of the system.  We have furthermore neglected the possibility of additional stars in the systems, which may affect the eccentricity in individual cases \citep{mazeh1979}. The probability of having third bodies is itself a (decreasing) function of orbital period \citep{tokovinin2006}. It may be interesting to analyze the shortest-period binaries with non-zero eccentricities, to see if they can be explained by the presence of a third companion.

Despite this long list of limitations, we have shown in a relatively direct and homogeneous manner that the eccentricity distribution of hot-hot and cool-cool binaries are significantly different as a function of orbital period. This is likely caused by a combination of the different age of the systems and a different tidal circularization efficiency. We anticipate that more detailed studied of (subsamples of) the \textit{Kepler} EB sample will further constrain tidal circularization theory. We also expect our findings to be of interest in the context of tidal theory for stellar obliquities in double star systems \citep[e.g.][]{albrecht2014} and in exoplanet systems \citep[e.g.][]{winn2010,albrecht2012}.\\

\acknowledgements

{\small We thank Kevin Schlaufman, Saul Rappaport, Jens Jessen-Hansen, and the anonymous referee for helpful comments and suggestions which have significantly improved this manuscript. Part of this manuscript was written at MIT and VVE appreciates the hospitality of the researchers and staff at the MIT Kavli Institute for Astrophysics and Space Research. Work by JNW was partly supported by funding from the NASA Origins program (grant ID NNX11AG85G). Funding for the Stellar Astrophysics Centre is provided by The Danish National Research Foundation (Grant agreement no.: DNRF106). The research is supported by the ASTERISK project (ASTERoseismic Investigations with SONG and Kepler) funded by the European Research Council (Grant agreement no.: 267864). We acknowledge ASK for covering travels in relation to this publication. Part of this work was supported by the Danish Council for Independent Research, through a DFF Sapere Aude Starting Grant nr.\ 4181-00487B.}

%\newpage
% \bibliographystyle{bibstyle}
% \bibliography{binaryecc_ref}

\end{document}